# Time Evolution of Quantum Fractals

Daniel Wójcik[1,2], Iwo Białynicki–Birula[1,3] and Karol Życzkowski[1,2,*]

[1] *Centrum Fizyki Teoretycznej, Polska Akademia Nauk, Al. Lotników 32/46, 02-668 Warszawa, Poland*
[2] *College of Science (Szkoła Nauk Ścisłych), Al. Lotników 32/46, 02-668 Warszawa, Poland*
[3] *Institute of Theoretical Physics, University of Warsaw, Poland*
*e-mail:* `danek@cft.edu.pl, birula@cft.edu.pl, karol@cft.edu.pl`

(September 20, 2000)

We propose a general construction of wave functions of arbitrary prescribed fractal dimension, for a wide class of quantum problems, including the infinite potential well, harmonic oscillator, linear potential and free particle. The box-counting dimension of the probability density $P_t(x) = |\Psi(x,t)|^2$ is shown not to change during the time evolution. We prove a universal relation $D_t = 1 + D_x/2$ linking the dimensions of space cross-sections $D_x$ and time cross-sections $D_t$ of the fractal quantum carpets.

Fractal objects are defined by their scaling behavior at the infinitely small or large scales. Such a requirement can be fulfilled by mathematical constructions (e.g. *Cantor set*), for which one assumes that the inductive algorithm is performed infinitely many times. Mathematical theory of fractals is well established [1,2].

In contrast, in the physical world any scaling property cannot hold for infinity of scales [1]. In spite of this obvious limitation, fractals found many applications in almost every branch of physics [1,3–6]. From the physicist's point of view, one tends to call an object *fractal*, if the scaling behavior can be observed for at least several orders of magnitude [1].

Can fractal behavior be found in quantum theory despite the coarse-graining effects of the Heisenberg uncertainty principle? M. V. Berry gave a partial answer to this question in an insightful paper [7]. He constructed a *quantum fractal in a box* from solutions of the Schrödinger equation for a particle in an infinite potential well. Assuming initial probability density to be uniform in the well, $|\Psi(\mathbf{x}, t=0)|^2 =$const, he showed that for (almost all) later times, $t > 0$, the probability function $P_t(\mathbf{x}) = |\Psi(\mathbf{x},t)|^2$ has a fractal nature and is characterized by the fractal dimension $D_x = D + 1/2 > 1$, where $D$ is the (Euclidean) dimension of the space. Berry's wave functions are initially *discontinuous* at the boundaries of the well. Due to propagation effects these initial discontinuities cause the wave function to become fractal.

In this Letter we propose a general construction of fractal solutions of the Schrödinger equation. Our construction scheme is valid for a large class of one, two, or three dimensional quantum problems. It can also be easily generalized to other wave phenomena in classical and quantum physics. For definiteness we demonstrate its usefulness for some textbook examples (infinite potential well, linear potential, harmonic oscillator and free particle). All these well known quantum problems are integrable, and our approach does not rely on *chaotic dynamics*. Moreover, our fractal wave functions are continuous everywhere.

Several quantum models related to chaotic scattering were also found to exhibit fractal-like structures [8,9] due to external perturbations or internal disorder. In our case fractality emerges in a regular system as a result of a special choice of the initial wave function.

We begin with the celebrated Weierstrass function [10,11]:

$$W(x) = \sum_{n=0}^{\infty} b^n \sin(a^n x), \quad a > 1 > b > 0, \quad ab \geq 1 \quad (1)$$

often quoted as an example of a continuous, nowhere differentiable function [2]. It exhibits fractal properties and the box dimension of its graph is $D_W = 2 - |\frac{\ln b}{\ln a}|$. Nonanaliticity of $W(x)$ is linked to the power-law behavior of its maximal oscillation $\mathrm{osc}_W$ on the interval of length $\tau$. It is known [2] that if $\mathrm{osc}_W(\tau) \sim \tau^\kappa$ for small $\tau$, where $\kappa$ is called the *Hölder exponent*, then the fractal (box counting) dimension is $\dim_B \mathrm{graph}[W(x)] = 2 - \kappa$. For the Weierstrass function $\mathrm{osc}_W(\tau) \sim \tau^{-\ln b/\ln a}$ [11].

Let us consider solutions of the Schrödinger equation $i\partial_t \Psi(x,t) = -\nabla^2 \Psi(x,t)$ for a particle in an infinite potential well. The general solutions satisfying the boundary conditions $\Psi(0,t) = 0 = \Psi(\pi, t)$ have the form

$$\Psi(x,t) = \sum_{n=1}^{\infty} a_n \sin(nx) e^{-in^2 t}. \quad (2)$$

In analogy to the Weierstrass function we construct fractal wave functions of the form

$$\Psi_M(x,t) = N \sum_{n=0}^{M} q^{n(s-2)} \sin(q^n x) e^{-iq^{2n}t}, \quad (3)$$

where $q = 2, 3, \ldots$, $2 > s > 0$.

In the physically interesting case of any finite $M$ the wave function $\Psi_M$ is a solution of the Schrödinger equation. The limiting case

$$\Psi(x,t) = \lim_{M \to \infty} \Psi_M(x,t), \quad (4)$$

with the normalization constant $N = \sqrt{\frac{2}{\pi}(1 - q^{2(s-2)})}$ is continuous but nowhere differentiable. It represents a vector in the Hilbert space which does not belong to the domain of the Hamiltonian $H$. However, it does belong to the domain of the unitary evolution operator. It can also be considered as a solution of the Schrödinger equation in the weak sense: for any fixed orthonormal basis $\{|\varphi_n\rangle\}$ for every $n$ we have

$$\langle H\varphi_n|\Psi(t)\rangle = i\hbar\frac{\partial}{\partial t}\langle \varphi_n|\Psi(t)\rangle. \qquad (5)$$

We show that not only the real part of the wave function $\Psi(x,t)$, but also the physically important probability density $P(x,t) = |\Psi(x,t)|^2$ exhibit fractal nature. This is not obvious, because $|\Psi(x,t)|^2$ is the sum of squares of real and imaginary part having usually equal dimensions. But the dimension of the graph of a sum of functions whose graphs have the same dimensions $D$ can be anything from 1 to $D$ (as seen from a simple example of $+f$ and $-f$).

In [12] we prove that the probability density $P(x,t)$ for the wave function (4) has the following properties:

1) *at the initial time* $t = 0$ *the probability density*, $P_0(x) = P(x,0)$, *forms a fractal graph in the space variable (i.e. a* space fractal*) of dimension* $D_x = \max\{s,1\}$;
2) *the dimension* $D_x$ *of graph of* $P_t(x)$ *does not change in time*;
3) *for almost every $x$ inside the well the probability density*, $P_x(t) = P(x = \text{const}, t)$, *forms a fractal graph in the time variable (i.e. a* time fractal*) of dimension* $D_t(x) = D_t := 1 + s/2$;
4) *for a discrete, dense set of points $x_d$, $P_{x_d}(t) = P(x_d,t)$ is smooth and thus $D_t(x_p) = 1$;*
5) *for even $q$ the average velocity $d\langle x\rangle/dt$ is fractal with its graph dimension equal to* $D_v = \max\{(1+s)/2, 1\}$;
6) *the surface $P(x,t)$ has the dimension $D_{xt} = 2 + s/2$.*

We will call the two-dimensional contour plots of the probability density $P(x,t)$ the *fractal quantum carpets* in analogy to the term *quantum carpets* used in the past [13]. In fig. 1 we show a typical fractal quantum carpet (lighter shade means greater probability) obtained for $q = 2$, $s = 3/2$, and its time and space cross-sections. Periodicity in time inhibits the decay of correlations. The period $2\pi/3$ visible in the carpet is connected with the structure of the frequency spectrum $\omega_{m,k} = 4^m - 4^{m-k} = 3(4^{m-1} + \ldots + 4^{m-k})$, $m = 1,\ldots,\infty$, $k = 1,\ldots,m$ of the probability density $P(x,t)$. The equal spacing between the clusters on the logarithmic scale

$$\ln \omega_{m,k} \approx m \ln 4 - 1/4^k \qquad (6)$$

allows for the fractal properties of $P(x,t)$ in time.

The dimensions $D_x$ and $D_t$ are the dimensions of the (generic) cross-sections of the carpet $P(x,t)$ in the space (Fig. 1c and 1d) and time (Fig. 1e) directions.

The Hölder exponent $\kappa$ in the $x$ direction is $(2-s)$ and in the $t$ direction is $(2-s)/2$ since the frequencies of time oscillations, $\exp(-iq^{2n}t)$, grow as $q^{2n}$. For these results to hold it is crucial that the series in equation (4) is infinite. This is not the case at the points $x_d = k\pi/q^m$ ($k = 0,1,\ldots,q^m$), for which the series (4) consists of at most $m$ terms and the time dependence $P_x(t)$ is smooth ($D_t = 1$). An example of this behavior is shown in figure 1f. Thus, $D_t(x) = \dim_B \text{graph}|\Psi(x = \text{const},t)|^2$ forms a nowhere continuous function on $[0,\pi]$; for almost all arguments $D_t(x) = D_t = 1 + s/2$, while it is equal to 1 for the dense set of points in this interval $\{k\pi/2^m\}$.

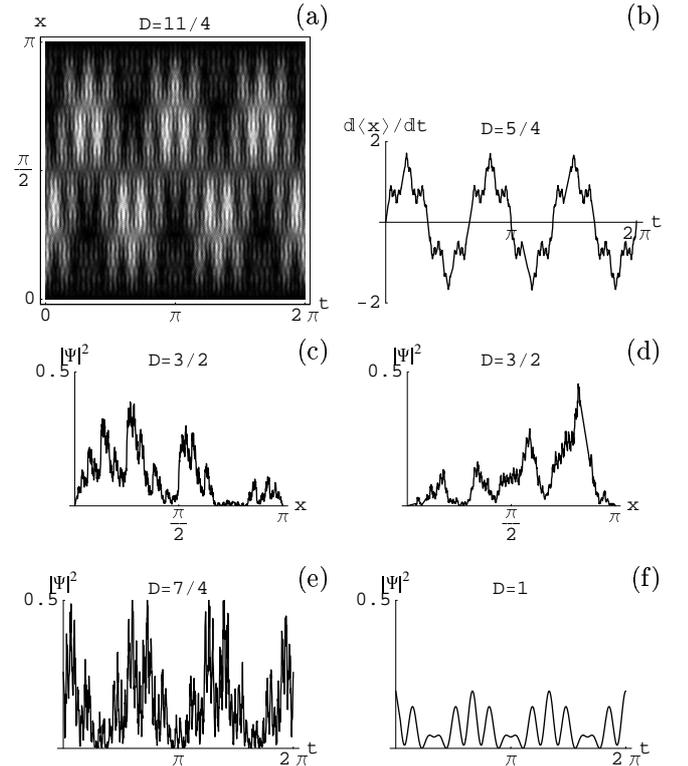

FIG. 1. Fractal quantum carpet in a box (a), average velocity (b), space (at $t = 0$ (c), and $t = 1$ (d)) and time (at $x = 1$ (e), and $x = \pi/8$ (f)) cross-sections obtained by superimposing $M = 20$ terms in eq. (3) with $q = 2$, $s = 3/2$.

Our fractal wave functions (4) have infinite mean energy necessary to generate mathematical fractals with infinite scaling. In practice a few terms in the series may lead to physically interesting effects. The lack of differentiability of $\Psi(x)$ leads to the average momentum $\langle p\rangle$ being ill-defined. On the other hand, we may compute the average position $\langle x\rangle$ and define the average velocity $v = d\langle x\rangle/dt$ *not* equal to $\langle p\rangle/m$ for infinite series (4). For physical fractals (3) we recover the Ehrenfest theorem and $\langle v\rangle = \langle p\rangle/m$. Straightforward calculations give a continuous function of the class $\mathcal{C}^1$, which for even $q$ is

$$\langle x\rangle = \frac{\pi}{2} - \frac{16(1-q^{2(s-2)})}{\pi}\sum_{k=1}^{\infty}\frac{q^{k(s-1)}\cos(q^{2k}-1)t}{(q^{2k}-1)^2}. \qquad (7)$$

It is a *semi-fractal*, since its first derivative $v$ produces a fractal graph with dimension $D_v = \max\{(1+s)/2, 1\}$ (Fig. 1b). For odd $q$, $\langle x \rangle = \pi/2$ and $D_v = 1$. In the region where all the three dimensions characterizing the fractal carpet $P(x,t)$ are non-integer ($2 > s > 1$), they are not independent but linked by the relation

$$D_t + D_v = D_x + 3/2. \quad (8)$$

Comprehensive analysis of the details of the fractal wave functions in the potential well is provided in [12].

Let us now demonstrate how to generate fractal solutions of the Schrödinger equation for the 1D harmonic oscillator. Let $\varphi_n(x)$ denote the eigenstates of this Hamiltonian expressed by the Hermite polynomials $H_n(x)$

$$\varphi_n(x) = (\sqrt{2\pi} 2^n n!)^{-1/2} H_n(x/\sqrt{2}) \exp(-x^2/4), \quad (9)$$

where $x$ is measured in units $\sqrt{\hbar/2m\omega}$.

Consider the following superposition of eigenstates

$$\Psi(x,t) = N \sum_{n=1}^{\infty} q^{n(s-3/2)} \varphi_{q^{2n}}(x) e^{-i(q^{2n}+1/2)t} \quad (10)$$

with a natural $q$ and a real $3/2 > s > 1$. On every fixed interval $[-x_0, x_0]$ sufficiently high energy eigenstates can be arbitrarily well approximated by

$$\varphi_n(x) \sim n^{-1/4} \sin[\sqrt{n+1/2}\, x - (n-1)\pi/2], \quad (11)$$

which follows from neglecting the potential in the WKB approximation. Thus the high $n$ contribution to the wavefunction (10) responsible for the fine scale structure of the probability density reduces to

$$\tilde{\Psi}(x,t) \approx \sum_{n=p}^{\infty} q^{n(s-2)} \sin(q^n x - \theta_{q^n}) e^{-i(q^{2n}+1/2)t}. \quad (12)$$

The analysis similar to the case of the infinite well allows one to obtain the dimensions of the cross-sections of the fractal carpet $|\Psi(x,t)|^2$

$$D_x = s \quad \text{and} \quad D_t = 1 + s/2 = 1 + D_x/2. \quad (13)$$

Note that this reasoning is valid for arbitrarily large $x_0$. However, the larger $x_0$ the more terms in the sum one needs to take to observe the fractal structure. An example of the fractal carpet for harmonic oscillator is provided in Fig. 2.

We conjecture that the relation $D_t = 1 + D_x/2$ is valid for a wide class of fractal solutions (not necessarily of the form (3)) of one-dimensional problems. This is due to the connection between the energy and momentum $E_n \sim k_n^2$ for high energy states (when one can neglect the potential). We provide here an explicit construction for asymptotically power-law potentials (for large $|x|$, $V(x) \sim |x|^\alpha$, where $\alpha > 0$). For these potentials we define the following superposition of the eigenstates:

$$\Psi_H(x,t) = N \sum_{n=1}^{\infty} q^{n(s-2+1/\alpha)} \varphi_{b_n}(x) e^{-iE_{b_n} t}, \quad (14)$$

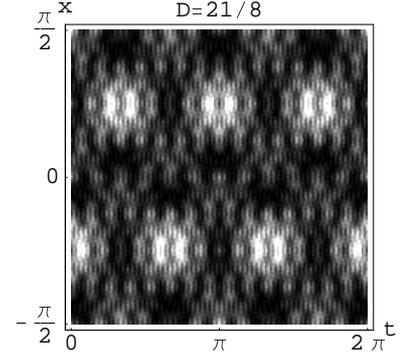

FIG. 2. Fractal quantum carpet $|\Psi(x,t)|^2$ of dimension $D_{xt} = 21/8$ for quantum harmonic oscillator obtained for $q=2$, $s=5/4$, and $M=15$ terms in (10).

where $b_n$ is the integer part of $q^{(1+2/\alpha)n}$, $q > 1$ and $2 > s > 0$. Such a function has the following properties:
1) is a normalizable solution of the corresponding Schrödinger equation in a weak sense for $s < 2 - 1/\alpha$,
2) for all $t_c$ the space dimension is constant, and $\dim_B \mathrm{graph}(P_{t_c}(x)) = D_x = \max\{1, s\}$,
3) for almost all $x_a$ the time dimension is constant, and $\dim_B \mathrm{graph}(P_{x_a}(t)) = D_t = 1 + s/2$,
4) for a countable dense set of $x_b$, $P(x_b, t)$ is differentiable and $\dim_B \mathrm{graph}(P_{x_b}(t)) = 1$;
5) the surface $P(x,t)$ has dimension $D_{xt} = 2 + s/2$.

To show this consider a superposition of eigenstates (14), by construction a weak solution of the Schrödinger equation. For large $n$ the WKB approximation gives eigenenergies $E_n \sim n^\beta$ with $\beta = 2\alpha/(2+\alpha)$ (e.g. [14]) and eigenstates $\varphi_n(x) \sim g_n \sin(k_n x - \theta_n)$, where the momentum $k_n = \sqrt{E_n} = n^{\beta/2}$. The normalization constant scales as $g_n = n^{-\gamma}$. To determine $\gamma$ consider the classical return points $x_n$ which satisfy: $n^\beta \sim E_n \sim x_n^\alpha$, thus $x_n \sim n^{\beta/\alpha} = n^{2/(2+\alpha)}$. Therefore

$$\int_{-x_n}^{x_n} g_n^2 \, dx \sim 1, \quad \text{so} \quad g_n \sim 1/\sqrt{x_n} = n^{-\gamma}, \quad (15)$$

where $\gamma = 1/(2+\alpha) = (2-\beta)/4$. Hence

$$\varphi_{B_n}(x) \sim g_{B_n} \sin(k_{B_n} x - \theta_{B_n}) \sim q^{-n/\alpha} \sin(q^n x), \quad (16)$$

and the wave function (14) might be written in the Weierstrass-like form

$$\Psi_H(x,t) \approx \sum_{n=1}^{\infty} q^{n(s-2)} \sin(q^n x - \theta_{B_n}) e^{-iq^{2n} t}. \quad (17)$$

Repeating the reasoning used earlier for the problem of rectangular well we conclude that if $2 - 1/\alpha > s > 0$ than for all $t$ the space dimension of the carpet is $D_x = \max\{1, s\}$, while for almost all $x_c$ the above series is infinite and the time dimension reads $D_t = 1 + s/2 = 1 + D_x/2$. This result can be generalized to $D$-dimensional power-law potentials for which $D_x = D + 2D_t - 3$ when $D_t \geq 3/2$ or $D_x = D$ otherwise [12].

| Generic | $\alpha$ | $\beta = 2\alpha/(2+\alpha)$ | $\gamma = 1/(2+\alpha)$ |
|---|---|---|---|
| Linear | 1 | 2/3 | 1/3 |
| Harm. Osc. | 2 | 1 | 1/4 |
| Well | $\infty$ | 2 | 0 |

TABLE I. The exponents $\alpha, \beta$ and $\gamma$ for the linear potential, harmonic oscillator and the infinite well.

The idea of producing fractal functions by taking a suitable superposition of eigenstates with the wavelengths of different orders of magnitude is very general and might be used for many problems. However, it cannot be easily generalized to continuous spectra. Let us consider, for example a free particle in one dimension. Following our constructions (eq. 3, 10, 14) we may write down a fractal superposition of the plane waves. To localize it we introduce a square integrable envelope $f(x)$. In this case the wave function is the sum of two terms of the form

$$\Psi_\pm(x,t) = \sum_n q^{n(s-2)} e^{\pm i q^n x - i q^{2n} t} f(x \mp 2q^n t, t), \quad (18)$$

where $f(x,t)$ is the solution of the Schrödinger equation satisfying $f(x,0) = f(x)$. The fractal properties of this packet will depend on the asymptotic behavior of $f(x,t)$ for large $x$. Fractality is destroyed if $f$ decreases sufficiently fast. This is the case for the Gaussian envelope of variance $\sigma^2$:

$$\Psi(x,t) = \frac{N_0}{\sqrt{1+it}} \exp\left[-\frac{x^2/4}{1+it}\right] \cdot \sum_{n=0}^\infty q^{n(s-2)} \sin\left[\frac{xq^n}{1+it}\right] \exp\left[\frac{-iq^{2n}t}{1+it}\right], \quad (19)$$

where $x$ is measured in units of $\sigma$ and $t$ in units of $2m\sigma^2/\hbar$.

For any non-zero value of $t$ the absolute value of the last term decreases exponentially with $n$ which destroys fractality. Therefore, this state (19) is fractal only at $t = 0$.

In conclusion, we have presented a general method of constructing fractal solutions of the Schrödinger equation and other equations that admit eigenfunction expansions. We have shown that fractals in quantum mechanics are equally legitimate as in the classical theory. In practice, the obvious physical limitations will not allow one to find in nature a Platonic structure of mathematical fractals, for which the property of scaling holds at all scales. On the other hand, one may well try to construct experimentally quantum states, for which scaling may be observed for at least a few decades. Since the number of energy levels of a particle of a mass $m$ in a potential well of characteristic length $L$ scales as $n \sim L\sqrt{m}$, the states described by the fractal wave functions might be easier to produce for heavy atoms, or ions, in macroscopic traps. Alternatively, one might look for the fractal solutions of the Maxwell equations in microwave cavities.

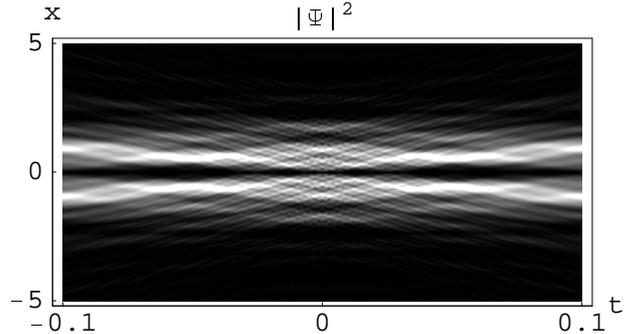

FIG. 3. Quantum carpet for free particle (19) with $M = 30$ terms, $s = 3/2$ and $q = 2$ displays fractal properties ($D_x = 3/2$) only at $t = 0$. For $t \neq 0$, $D_x = 1$.

It is a pleasure to thank P. Garbaczewski, S. Gnutzmann, J. Kijowski and M. Robnik for valuable remarks. Financial support by a research grant of Komitet Badań Naukowych in Warsaw is gratefully acknowledged.